\documentclass[aps,rmp,twocolumn,american]{revtex4}
\usepackage{amsmath,amssymb,amsxtra,amsfonts}
\bibpunct{[}{]}{;}{,}{,}

\makeatletter

\providecommand{\LyX}{L\kern-.1667em\lower.25em\hbox{Y}\kern-.125emX\@}

\usepackage{pslatex}
\usepackage{citesort}
\makeatother
\begin{document}

\title{\Huge What's Wrong with this Rebuttal?\footnote{Appeared as: {\it Found.
Phys. Lett.} {\bf 19} (6) 625-629 (2006).}}

\author{\textbf A. F. Kracklauer}

\homepage{www.nonloco-physics.000freehosting.com}

%\address{\emph{Belvederer Allee 23c; 99425 Weimar, Germany}}

%\email{\emph{kracklau@fossi.uni-weimar.de}}

\begin{abstract}
A recent rebuttal to criticism of Bell's analysis is shown to be defective
by fault of failure to consider all hypothetical conditions input
into the derivation of Bell Inequalities.
\end{abstract}
\maketitle

\section{The dispute}

On the pages of \emph{Found. Phys.} \cite{key-1}, under the title:
``What's wrong with this Criticism?'', N. D. Mermin rebutted criticism
of Bell's Theorem and analysis by K. Hess and W. Philipp \cite{key-2}.
The latter authors hold that Bell's analysis of EPR inspired experiments
testing the contention that Quantum Mechanics could be `completed,'
i.e., rendered a deterministic theory instead of only a probabilistic
one, is fatally flawed. They argue that Bell failed to consider time
variable correlations and subsequently failed to find structure permitting
a local realistic interpretation of the results of EPR experiments.

While this is the context of the larger dispute, the actual point
of contention for Mermin was a much narrower, although equally potent,
sub-argument that Hess and Philipp mentioned along the way. It is
essentially this: For technical reasons, data taken in feasible experiments
cannot meet all requirements in the input into derivations of Bell
Inequalities. This is an old observation, although its many renditions
are not always easily recognized as being the same issue.

To crystallize the crucial points, recall that Bell's analysis ostensibly
proves that for all local realistic theories a certain expression,
in Mermin's notation denoted by $\Gamma $, satisfies: 
\begin{equation}
|\Gamma |\leqslant 2.\label{eq:1}
\end{equation}
Now $\Gamma $, as is easily seen from its derivation (which is very
well known and will not be reiterated here; see: \cite{key-3}), for EPR-type
experiments is comprised of a particular sum of terms, where each
one is the sum of the products of the outcomes in each arm for a given
combination of the polarizer settings (or Stern-Gerlach field directions
if the experimental objects are electrons). In so far as two different
settings are considered for each arm, there are then four combinations
so that $\Gamma $ can be written as: 
\begin{equation}
\Gamma ={1 \over N}\sum _{j}[a_{\textrm{a}}(j)b_{\textrm{b}}(j)+
a_{\textrm{a}}(j)b_{\textrm{c}}(j)+ a_{\textrm{d}}(j)b_{\textrm{b}}(j)-
a_{\textrm{d}}(j)b_{\textrm{c}}(j)].\label{eq:2}
\end{equation}

It is just here that Hess and Philipp, as have others before them%
\footnote{Recent studies by Adenier \cite{key-9} and Sica \cite{key-10} explictly
make the same point. Apparently, de la Pe\~na, Cetto and Brody \cite{key-11}
were first to recognize the significance of the relevant structure.%
}, raise an objection. It is that for the different settings of the
polarizers, that is, for the various of the four combinations or each
term alone, one has essentially four different experiments and that
it is not legitimate to mix the data and analyze it as if it came
from a single experiment.

\section{Mermin's contention}

In response, Mermin's rebuttal consists of asserting that Eq. (\ref{eq:2})
can also be written:
\begin{align}
\Gamma  & =(1/N_{\textrm{ab}})\sum _{j\in
X_{\textrm{ab}}}a_{\textrm{a}}(j)b_{\textrm{b}}(j)+(1/N_{\textrm{ac}})\sum
_{j\in X_{\textrm{ac}}}a_{\textrm{a}}(j)b_{\textrm{c}}(j)+\nonumber \\ + &
(1/N_{\textrm{db}})\sum _{j\in
X_{\textrm{db}}}a_{\textrm{d}}(j)b_{\textrm{b}}(j)-(1/N_{\textrm{dc}})\sum
_{j\in X_{\textrm{dc}}}a_{\textrm{d}}(j)b_{\textrm{c}}(j)],\label{eq:3}
\end{align}
and that, in a sufficiently long run, that is large enough $N$, ``each
of the four choices for $xy$ the $N_{xy}$ indices $j$ appearing
in $X_{xy}$ constitute a random sample of the full set $j=1\ldots N$,
each $j$ having the probability $1/4$ of appearing in $X_{xy}$.
So by standard sampling theory ...''

\section{A lacuna in the rebuttal}

The point of this rebuttal is necessary but not sufficient, however.
The data collected in four equal length sub-runs, one each for each
polarizer combination, must satisfy \emph{all} the hypothetical inputs
into the derivation of a Bell Inequality, e.g., Eq. (\ref{eq:1}).
That is, the factor sequence $a_{\textrm{a}}(j)$ in the first term
of Eq. (\ref{eq:3}) must be identical with the same factor sequence
in the second term. Specifically, this means that if in, e.g., $a_{\textrm{a}}(j)$,
`$+1$' appears $k$ times, then not only must `$+1$' appear $k$
times in $a_{\textrm{a}}(j)$ in the second term (that it does indeed,
is Mermin's point), but additionally, and neglected in Mermin's rebuttal,
the pattern of switches back and forth among the `$+1$' and `$-1$'s'
must occur at the same locations along the sequence. This is, it turns
out, necessary because it is an implicit assumption in the derivation
of Bell Inequalities (regardless of the detailed logic and discriminator. i.e.,
$\Gamma$,
for any particular derivation of an inequality). It results from steps
in the derivation employing factoring among the factor sequences.
In random runs of the experiment, it can be expected that the ratio
of `$+1$'s' to `$-1$'s' will be equal for sufficiently long subsequences
with fixed polarizer settings, just as Mermin surmises; but their
pattern of occurrence will not repeat---indeed, this is what is meant
by being ``random'' in this case. Moreover, the correlation existing
between the outcomes on both sides of an EPR experiment (one of Bell's
hypothetical inputs is that the pairs are correlated) implies that
the pattern of matches between the two sides in general is not fully
random---it follows an extention of Malus' Law, in fact. 

We shall not go into all the details here as this matter has been
explicated in detail elsewhere \cite{key-5,key-8}. But in conclusion,
we see that the ``very simple way past {[}Hess' and Philipp's{]}
objection'' does not lead to the conclusion that Mermin wishes to
support.

\section{The structure illustrated}

To further illustrate the logical impossibility of relating data taken
from feasible experiments with sequences required for Eq. (\ref{eq:3}),
consider the following.

First, let us simplify notation as follows: \begin{equation}
(1/N_{\textrm{ab}})\sum _{j\in X_{\textrm{ab}}}a_{\textrm{a}}(j)b_{\textrm{b}}(j)=<a_{1}b_{1}>,\label{eq:4}\end{equation}
so that Eq. (\ref{eq:3}) becomes\begin{equation}
\Gamma =<a_{1}b_{1}>+<a_{2}c_{2}>+<d_{3}b_{3}>-<d_{4}c_{4}>.\label{eq:5}\end{equation}

Now, if $a_{1}\equiv a_{2}$, $b_{1}\equiv b_{3}$, $c_{2}\equiv c_{4}$
and $d_{3}\equiv d_{4}$ (here identity means both that the sequences
have the same quantity of $+1$'s and the same pattern of switches
between $+1$'s and $-1$'s), then Eq. (\ref{eq:5}) can be written:\begin{equation}
\Gamma =<a(b+c)>+<d(b-c)>,\label{eq:6}\end{equation}
so that in the sum over all $j$ the absolute value of either one
term or the other is $2$ while its partner is $0$ alternatively,
such that the average satisfies Eq. (\ref{eq:1}).%
\footnote{The termwise factorizations leading to Eq. (\ref{eq:6}) are parallel
to those in the derivation of a Bell Inequality; in both cases factorization
requires identically ordered sequences in order to be executable.%
} This in turn leads one to imagine that if $a_{2}$ is re-sorted to
have the identical pattern of switches as $a_{1}$, then perhaps the
data taken from a feasible experiment can be re-sorted so as to satisfy
\emph{all} the hypothetical inputs into the derivations of Bell Inequalities.

This turns out not to be possible, however. Suppose the second term
in Eq. (\ref{eq:5}) is re-sorted so that $\tilde{a_{2}}$ (the tilde
indicates the re-sorted variant) is essentially identical with $a_{1}$,
then the second re-sorted term becomes $<a_{1}\tilde{c_{2}}>$. This
new ordering must be cascaded to the fourth term to become $<\tilde{d_{4}}\tilde{c_{2}}>$,
and finally to the third term: $<\tilde{d_{4}}\tilde{b_{3}}>$. The
final result is that Eq. (\ref{eq:6}) becomes: 
\begin{equation}
\Gamma=<a_{1}(\tilde{b_{1}}+\tilde{c_{2}})> +
<\tilde{d_{4}}(\tilde{b_{3}}-\tilde{c_{2}})>.\label{eq:7}
\end{equation}
In order for $|\Gamma |$ in this form always to be less than or equal
to $2$, it is necessary that $\tilde{b_{1}}\equiv \tilde{b_{3}}$;
but here, separate independent re-sortings are involved; they would
be essentially identical only under the most improbable circumstances.
The circuit cannot be closed; there is no reason from the physics
of the situation, classical by explicit supposition, why these separate
re-sortings must be the same. As a consequence, we see, data from feasible
experiments cannot be re-sorted to comply with all the hypothetical
inputs into derivations of Bell inequalities, and therefore, they
cannot be used in Eq. (\ref{eq:1}) to discriminate between local-realistic
and other theories.

\section{Quo vadimus?}

These `re-sorting' considerations expose exactly the structure that
torpedoes Mermin's rebuttal of Hess' \& Philipp's criticism of Bell's
analysis. At the same time, however, they cannot be turned around
to support Hess' \& Philipp's particular analysis critical of Bell's
arguments. Indeed, this writer argues elsewhere \cite{key-5}, that
time variable correlations are not necessary to invalidate the logic
of Bell's approach. A much more incisive line of fatal critical analysis
was founded by Edwin Jaynes \cite{key-6}%
\footnote{The arguments made by Hess and Philipp and by Jaynes are to be distinguished
from others usually designated ``loopholes.'' The latter concern
only features rendering experiments ambiguous or invalid; the former
concern fundamental defects that pertain even when detectors are 100\%
efficient, etc.%
}. He, apparently, was first to observe that the probabilistic nature
of EPR and GHZ experiments requires Bayes' formula for \emph{conditional}
probabilities, which, if introduced into Bell-type analysis, prevents
the derivation altogether of inequalities for the very same structural
reasons reviewed above and thus far overlooked by proponents of Bell's
``Theorem.'' (For details see, e.g., \cite{key-7}.) 

\bibliography{unsrt}

\end{document}